\documentclass[journal]{IEEEtai}

\usepackage[colorlinks,urlcolor=blue,linkcolor=blue,citecolor=blue]{hyperref}

\usepackage{color,array}

\usepackage{graphicx}
\usepackage{cite}
\usepackage{amsmath,amssymb,amsfonts}
\usepackage{algorithmic}
\usepackage{color,array}
\usepackage{graphicx}
\usepackage{array}
\usepackage{acronym}
\usepackage{hyperref}
\usepackage{textcomp}
\usepackage{multirow}

\acrodef{ADM}[ADM]{asynchronous  delta modulator}
\acrodef{ECG}[ECG]{electrocardiography}
\acrodef{EMG}[EMG]{electromyography}
\acrodef{EEG}[EEG]{electroencephalography}
\acrodef{ECoG}[ECoG]{electrocorticography}
\acrodef{ENG}[ENG]{electroneurographic}
\acrodef{HFO}[HFO]{High-Frequency Oscillations}
\acrodef{LIF}[LIF]{Leaky Integrate and Fire}
\acrodef{LIFE}[LIFE]{Longitudinal Intrafascicular Electrode}
\acrodef{TIME}[TIME]{Transverse Intrafascicular Multichannel Electrode}
\acrodef{tfLIFE}[tf-LIFE]{thin-film LIFE} 
\acrodef{DoF}[DoF]{degree-of-freedom}
\acrodefplural{DoF}{degrees-of-freedom}
\acrodef{MAV}[MAV]{mean absolute value} 
\acrodef{SNN}[SNN]{Spiking Neural Network}
\acrodef{USEA}[USEA]{Utah slanted electrode array}
\acrodef{RBF}[RBF]{radial basis function}
\acrodef{RMS}[RMS]{root mean square}
\acrodef{SVM}[SVM]{Support Vector Machine}
\acrodef{LDA}[LDA]{Linear Discriminant Analysis}
\acrodef{ANN}[ANN]{artificial neural network}
\acrodef{PCA}[PCA]{Principal Component Analysis}
\acrodef{PC}[PC]{Principal Component}
\acrodef{CCA}[CCA]{Canonical Correlation Analysis}
\acrodef{VLSI}[VLSI]{Very Large Scale Integration}
\acrodef{PRC}[PRC]{physical reservoir computing}
\acrodef{RNR}[RNR]{Rotating neuron reservoir} 
\acrodef{sRNR}[sRNR]{spiking RNR}
\acrodef{sEMG}[sEMG]{Surface electromyography}
\acrodef{k-NN}[k-NN]{k-Nearest Neighbours} 
\acrodef{LDA}[LDA]{Linear Discriminant Analysis} 
\acrodef{RF}[RF]{Random Forests}
\acrodef{ReLU}[ReLU]{rectified linear units}
\acrodef{eRNR}[eRNR]{electronic RNR} 
\acrodef{ADC}[ADC]{analog-to-digital converter}
\acrodef{ODE}[ODE]{ordinary differential equation}
\acrodef{CE}[CE]{cross-entropy}
\acrodef{t-SNE}[t-SNE]{t-distributed Stochastic Neighbor Embedding}
\acrodef{RNN}[RNN]{Recurrent Neural Network}
\acrodef{CNN}[CNN]{Convolutional Neural Network}
\acrodef{DNN}[DNN]{Deep Neural Network}
\acrodef{ALR}[ALR]{Auto-Labelling-Refining}
\acrodef{RC}[RC]{Reservoir Computing}
\acrodef{WTA}[WTA]{Winner-Take-ALL}
\acrodef{TD}[TD]{Time-Domain}
\acrodef{HIST}[HIST]{Histogram}
\acrodef{mDWT}[mDWT]{marginal Discrete Wavelet Transform}
\acrodef{STDP}[STDP]{Spiking-Timing Dependent Plasticity}
\acrodef{sFCN}[sFCN]{spiking fully connected layer}
\acrodef{PP}[PP]{Positive Predictivity}
\acrodef{Sp}[Sp]{specificity}
\acrodef{Se}[Se]{Sensitivity}
\acrodef{TP}[TP]{True Positive}
\acrodef{TN}[TN]{True Negtive}
\acrodef{FP}[FP]{False Positive}
\acrodef{FN}[FN]{False Negative}
\begin{document}

\title{Event-Driven Implementation of a Physical Reservoir Computing Framework for superficial EMG-based Gesture Recognition}

\author{Yuqi Ding, \IEEEmembership{Student Member, IEEE}, Elisa Donati, \IEEEmembership{Member, IEEE}, Haobo Li, \IEEEmembership{Member, IEEE} and Hadi Heidari, \IEEEmembership{Senior Member, IEEE}
\thanks{This work was partially supported by the United Kingdom EPSRC under grants EP/X031950/1 and EP/X034690/1. (Corresponding author: Hadi Heidari.)}
\thanks{Yuqi Ding and Hadi Heidari are with the James Watt School of Engineering, University of Glasgow, Glasgow, G12 8QQ, UK (email: y.ding.4@research.gla.ac.uk and Hadi.Heidari@glasgow.ac.uk). }
\thanks{Elisa Donati is with the Institute of Neuroinformatics, University of Zurich and ETH Zurich, 8057 Zurich, CH (e-mail: elisa@ini.uzh.ch).}
\thanks{Haobo Li was with University of Glasgow, Glasgow, G12 8QQ, UK. He is now with the School of Science and Engineering, University of Dundee, Dundee, DD1 4HN, UK (e-mail: hli005@dundee.ac.uk).}
\thanks{This paper is a preprint submitted to arXiv.}
}


\maketitle

\begin{abstract}
Wearable health devices have a strong demand in real-time biomedical signal processing. However traditional methods often require data transmission to centralized processing unit with substantial computational resources after collecting it from edge devices. Neuromorphic computing is an emerging field that seeks to design specialized hardware for computing systems inspired by the structure, function, and dynamics of the human brain, offering significant advantages in latency and power consumption. This paper explores a novel neuromorphic implementation approach for gesture recognition by extracting spatiotemporal spiking information from surface electromyography (sEMG) data in an event-driven manner. At the same time, the network was designed by implementing a simple-structured and hardware-friendly Physical Reservoir Computing (PRC) framework called Rotating Neuron Reservoir (RNR) within the domain of Spiking neural network (SNN). The spiking RNR (sRNR) is promising to pipeline an innovative solution to compact embedded wearable systems, enabling low-latency, real-time processing directly at the sensor level. The proposed system was validated by an open-access large-scale sEMG database and achieved an average classification accuracy of 74.6\% and 80.3\% using a classical machine learning classifier and a delta learning rule algorithm respectively. While the delta learning rule could be fully spiking and implementable on neuromorphic chips, the proposed gesture recognition system demonstrates the potential for near-sensor low-latency processing.
\end{abstract}

\begin{IEEEImpStatement}
While deep learning dominates sEMG-based gesture recognition, its high computational cost limits real-time, low-power applications. Neuromorphic computing offers an energy-efficient alternative, making it ideal for edge computing in resource-constrained environments. This work is the first to integrate a PRC framework, specifically the RNR, within an SNN architecture and introduce a novel event-based encoding scheme to convert superficial EMG signals into spike trains. Our approach surpasses existing SNN-based methods in classification accuracy while remaining competitive with deep learning models in a more lightweight form. Crucially, the use of recurrent reservoirs addresses a fundamental challenge in neuromorphic systems—the absence of built-in memory. By generating memory at the network level, this method enables robust, real-time processing of dynamic signals, which is essential for real-world biomedical applications. This advancement paves the way for next-generation wearable systems with ultra-low latency and embedded intelligence.
\end{IEEEImpStatement}


\section{Introduction}
\label{sec:intro}

\begin{figure*}[htp]
\centerline{\includegraphics[width=\textwidth]{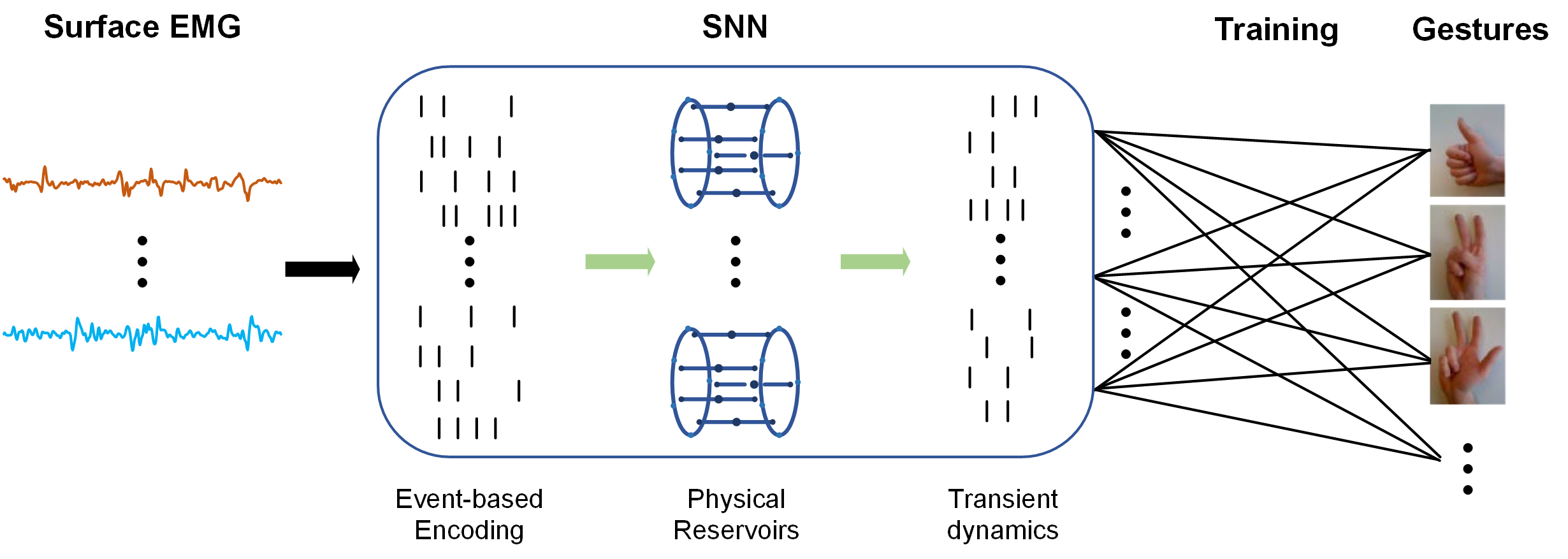}}
\caption {Architecture of the proposed classification system. The raw sEMG signals are encoded into SNN-compatible spike trains by an event-based encoding scheme. An SNN consisting of physical reservoirs is used to generate transient responses to a higher dimensional feature space. The collected dynamical states are trained in the readout layer only by machine learning algorithms for classifying gestures.}
\label{fig:conceptual}
\end{figure*}

\ac{sEMG} is a non-invasive technique that reflects the electrical activities of skeletal muscle movements, providing insights into muscle cells essential for diagnosis, rehabilitation and other medical applications. \ac{sEMG}-based gesture classification is a specialized application that focuses on recognizing motor intentions by analyzing the electrical activity of muscles. This field enables applications involving wearable devices in areas such as human-computer interaction, prosthetics control and robotics~\cite{guo2021human,donati2019discrimination}. Processing data locally on the wearable device or nearby, rather than sending data to a centralized server is crucial to perform real-time analysis, save bandwidth, improve privacy and save battery. To further improve local processing, it is crucial to identify lightweight gesture recognition algorithms. Traditional machine learning approaches, such as, \ac{SVM}, \ac{k-NN}, \ac{LDA}, and \ac{RF} require feature extractions from both the time domain and frequency domain~\cite{atzori2014electromyography}. Deep learning models have recently achieved higher classification accuracy on raw \ac{sEMG} signals by automatically learning features~\cite{wei2019surface,fatayer2022semg}. However, this comes at the cost of increased power consumption and memory requirements.

Neuromorphic computing, which prioritizes low latency and energy efficiency, is another promising approach for processing temporal signals at the edge~\cite{schuman2017survey,azghadi2020hardware, covi2021adaptive, donati2023neuromorphic}. Among neuromorphic systems, \ac{PRC} has emerged as a compelling solution. \ac{PRC} harnesses the intrinsic dynamics of physical systems with collocated memory and processing units, distinct from traditional Von Neumann architecture, to accelerate and reduce the memory requirements of machine learning computations. This approach aligns well with the demands of edge computing, where information processing is performed closer to sensors, minimizing the latency associated with data transmission~\cite{nakajima2020physical,liang2024physical,wright2022deep}. At the same time, \ac{RC} is a form of \acp{RNN} used primarily for processing time-series data, but what makes it different is the use of a fixed, untrained network called a "reservoir" that projects input data into a high-dimensional space. In this high-dimensional space, complex temporal patterns can be more easily processed with training readout layer only to avoid costly operations for backpropogation associated \acp{DNN}. 

\ac{RNR} is a novel \ac{PRC} paradigm characterized by its simple, hardware-compatible architecture, realized through an analog electronic circuit~\cite{liang2022rotating}. This architecture has demonstrated its efficacy in various temporal signal forecasting applications~\cite{ding2022mmg,ding2024physical}. For the first time we propose an \ac{SNN} based \ac{RNR} framework, more biologically plausible than traditional \acp{ANN} based on sigmoid units or \ac{ReLU}~\cite{yamazaki2022spiking,maass1997networks,roy2019towards}. This incorporation offers three key benefits: (i) A spike is a single-bit event, either a '1' or a '0', which is more hardware-friendly than floating point values. In comparison, the \ac{eRNR}~\cite{liang2022rotating} where signals are processed in the analog domain. However, floating point values precision in analog circuits is costly in terms of complexity and hardware costs~\cite{kendall2020building}, hence, processing information in the form of low-precision spike trains could be a possible solution. (ii) event-driven processing allows for energy efficiency and low latency as the neurons only respond when an event occurs, leading to sparse vectors/tensors that are cheap to store and low-power to move~\cite{tavanaei2019deepSNN}. (iii) This fusion still retains the advantages of \ac{RNR}'s simple and hardware-friendly structure with training readout layer only. At the same time, the low-precision reservoir states enables the use of trainable classifiers to further improve classification performance.

Encoding \ac{sEMG} signals into spatiotemporal spiking information plays an important role in executing \acp{SNN}. A common signal-to-spike encoding technique is the delta-modulator analog-to-digital converter~\cite{donati2019discrimination,ma2020emg,vitale2022neuromorphic}. However, the classification performance by using this strategy lags behind the state-of-the-art, especially compared with deep learning methods. In our work, we adopted an event-based encoding fashion for \ac{sEMG} signals which was validated by a regression task. This method is inspired by how mammalian cochlea processes auditory signals and has been applied for feature extraction of neural and audio signals~\cite{corradi2015neuromorphic,yang20160}. It extracted enough informative features from \ac{sEMG} signals and performed well in a force estimation regression task~\cite{zanghieri2024eventencoding}. The integration of this encoding scheme and our proposed \ac{sRNR} network escalated the \ac{SNN}-based gesture recognition accuracy to a new level.

The proposed gesture recognition system was verified by an open-access database. While the use of \acp{SNN} for \ac{sEMG}-based gesture recognition has been explored previously, our research offers substantial advancements in this field, as detailed below:
\begin{itemize}
\item We incorporated a \ac{PRC} framework called \ac{RNR} with an \ac{SNN} scheme for the first time to achieve sparse and low-precision data representation beneficial to memory requirements and low latency. While \ac{PRC} generates dynamics directly through physical systems beneficial to resource-efficient information processing, the \ac{RNR} paradigm additionally outperforms other \ac{PRC} paradigms by offering a simplified hardware design that is more easily interpretable by algorithms. This design minimizes the need for modules such as \ac{ADC}, buffer and memory, thereby reducing overall system complexity.

\item The proposed network has a fixed and simplified topology which is hardware-friendly and capable of using trainable nonlinear classifiers. Additionally, only the readout layer requires training, resulting in fewer parameters monitoring and reduced operations compared with backprogation-aided \acp{DNN}.

\item The proposed work obtained state-of-the-art performance among \acp{SNN} and demonstrated competitiveness with deep learning methods.
\end{itemize}

\section{Methods}
\label{sec:methods}

We evaluated our method using a publicly available \ac{sEMG} dataset. Encoding signal to spikes is a crucial step in event-based signal processing~\cite{bian2024evaluation}. In this work, the signals were converted into spike trains using an event-based encoding scheme inspired by the mammalian cochlea~\cite{zanghieri2023event, zanghieri2024eventencoding} to feed into a novel \ac{RNR}, implemented in a \ac{SNN} using SNNtorch~\cite{eshraghian2021SNNtorch} to classify hand gestures.

\begin{figure*}[!t]
\centerline{\includegraphics[width=\textwidth]{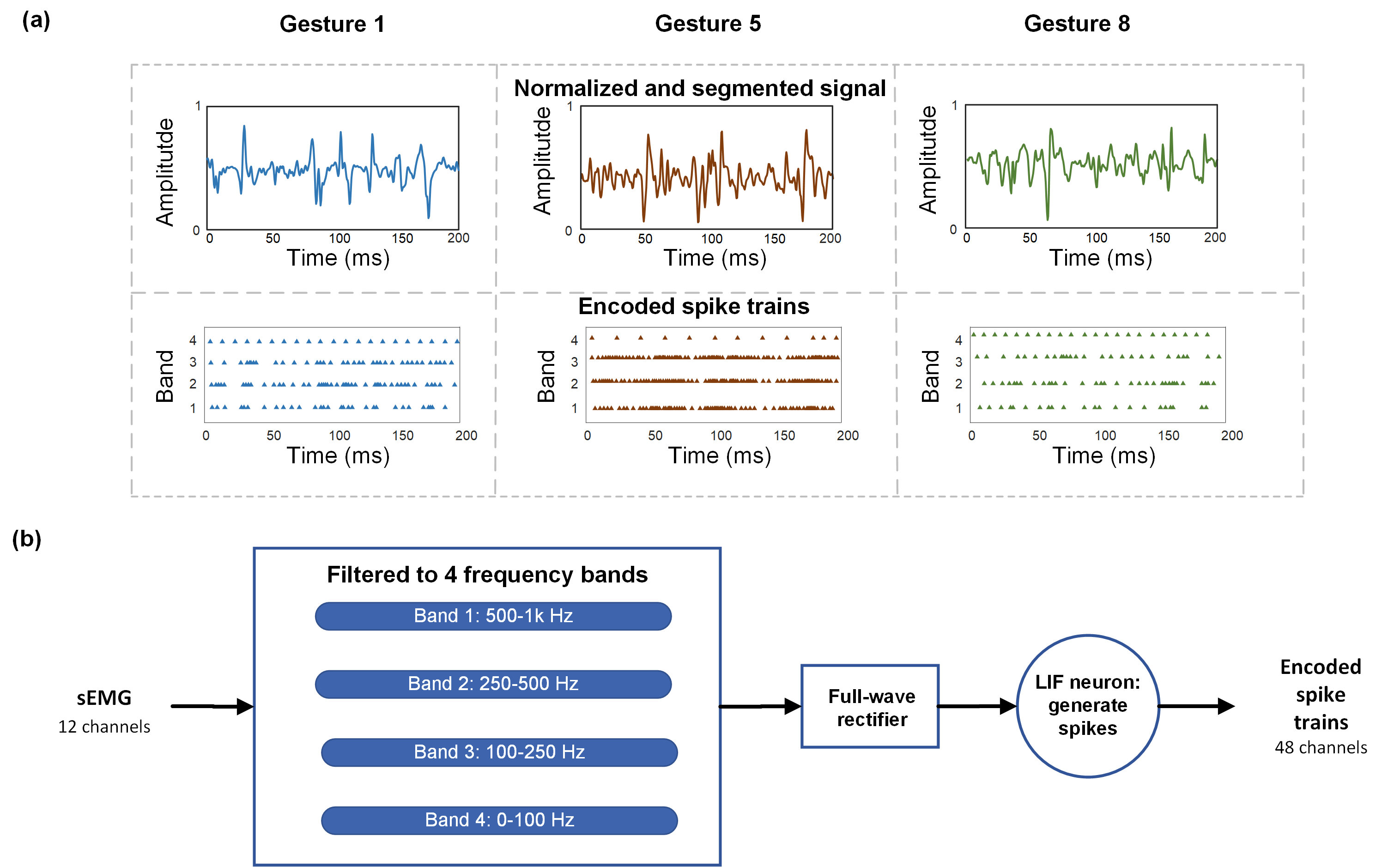}}
\caption {(a) Examples of one channel of normalized sEMG signals and encoded spike trains for Gesture 1, Gesture 5 and Gesture 8 respectively. (b) The process of spike encoding. After encoding, the original 12 channels of sEMG signals are encoded to 48 channels of spike trains. }
\label{fig:energy-based}
\end{figure*}

\subsection{Dataset Description and Preprocessing}
\label{ssec:dataset}
NinaPro (Non-Invasive Adaptive Prosthetics) databases are open-access \ac{EMG} datasets that are commonly used in hand gesture classification tasks~\cite{atzori2014electromyography}. In our project, the NinaPro DB2 was chosen to validate the proposed approach. This database is composed of 40 intact subjects (28 males, 12 females) performing 50 gestures with 6 repetitions for each. The \ac{sEMG} signals were collected by 12 Delsys Trigno Wireless electrodes placed on the forearm and sampled at a rate of 2 kHz. Each movement repetition lasts 5 s and then returns to the rest state for 3 s to remove any residual muscular activation. The 50 gestures include exercise B, C, D and rest position: exercise B comprises 8 isometric and isotonic hand configurations and 9 wrist movements; exercise C comprises 23 grasping movements which are common daily-life actions; exercise D includes 9 force patterns~\cite{atzori2014electromyography}.

Before converting the signal to spike trains that are compatible with an \ac{SNN} architecture, a pre-processing procedure including normalization and segmentation is required to construct the training and testing datasets, as described in the following steps:
\begin{enumerate}
\item \textit{Normalization:} To make it easier to compare \ac{sEMG} signals from different repetitions and subjects and ensure that all signals have the same data distributions, a normalization procedure is applied to scale the signals to the range of (0,1) by using the following equation:
\begin{equation}
signal_{normalized} = \frac{signal - min(signal)}{max(signal)-min(signal)}
\label{eq:eq1}
\end{equation}

\item \textit{Segmentation and Windowing:} In biomedical recognition tasks, the acquisition of precisely labeled datasets is essential. As some participants may start the movement earlier than the actual motion and some may start later, we only investigated the steady state of the movements. Our work omitted the initial and final 600 ms of each repetition, which may include gesture transition states, similar to the work done in~\cite{ma2020emg} for intuitive comparison. Besides, an optimal window length of 150-250 ms is preferred regarding classification performance latency in prosthetic control~\cite{Smith2011windowlenth}. Given the constraints, we constructed a dataset with an equal distribution of gestures, employing a window length of 200 ms for each segment. The examples of preprocessed \ac{sEMG} signals with a window length of 200 ms can be viewed in Fig.~\ref{fig:energy-based}a.

\item \textit{Training and Testing Sets:} We focused on evaluating the intra-subject gesture classification performance. For each subject, the training and testing sets were shuffled and split by a 4:1 ratio. The classification accuracy and standard deviation were obtained by averaging over all 40 subjects included in Ninapro DB2.
\end{enumerate}

\subsection{Spike encoding}
\label{ssec:encoding}
Encoding the raw \ac{sEMG} into spike trains is a crucial step in terms of performing spike-based gesture classification. The spike sequences should convey as much intrinsic information about muscle activity included in the raw \ac{sEMG} signals as possible during the encoding process. Inspired by how mammalian cochlea process auditory signals, we adopted an event-based encoding scheme in our project that exploits the raw \ac{sEMG} signals to four different frequency bands and encodes the extracted signals in each frequency band to spike trains by \ac{LIF} neurons based on the energy of the frequencies~\cite{zanghieri2023event, zanghieri2024eventencoding}. Since in this article we performed a classification task, the spike encoding is slightly different. The detailed encoding strategy is explained below and a flow chart of this process is illustrated in Fig.~\ref{fig:energy-based}b.  

\begin{enumerate}
\item \textit{Bandpass Filtering: } Each channel of the raw \ac{sEMG} signals is filtered by a four 4th-order Butterworth bandpass filter. According to the Nyquist-Shannon Sampling Theorem, 
as the sampling frequency for NinaPro DB2 is $f_{sampling}=2 kHz$, the max cutoff frequency for the frequency bands should be $f_{max} = 1 kHz$. Besides, the energy of \ac{sEMG} signals is mainly concentrated within the frequency range of 10 to 500Hz~\cite{li2020review}. In this case, we split the (0,1 kHz) band into 4 frequency bands, denoted by the logarithmic-distributed cutoff frequencies presented in equation (2). 
\begin{equation}
f_{cutoff} = \{0, 100, 250, 500, 1000\}Hz
\label{eq:eq2}
\end{equation}
The 12 channels of raw \ac{sEMG} signals will be expanded to 48 channels accordingly.

\item \textit{Full-wave rectifying: } The injected currents in \ac{LIF} neurons are required to be positive, therefore, a full-wave rectifier is applied to keep all components positive.

\item \textit{Genereating spikes through \ac{LIF} neurons: } The \ac{LIF} neuron is a simplified model used to describe the electrical characteristics of a biological neuron. The dynamics of the \ac{LIF} neuron is governed by the following differential equation (3):
\begin{equation}
\frac{dV(t)}{dt} = \frac{-(V(t)-V_{rest})}{\tau} + \frac{I(t)}{C_m}
\label{eq:eq3}
\end{equation}
where $\tau$ is the time constant for membrane relaxation time, $V(t)$ is the membrane potential at time $t$, $V_{rest}$ is the baseline membrane potential when the neuron is inactive, $I(t)$ is the injected current that drives the membrane potential and $C_m$ is the membrane capacitance. $S_{out}(t) \in \{0,1\}$ is the output spike train generated by the neuron. The neuron will fire a spike ($S_{out}(t) = 1$) when the membrane potential $V(t)$ exceeds the threshold $V_{thr}$ and it is reset to $V_{reset}$. Otherwise, the reset term will not be applied ($S_{out}(t) = 0$). This process can be denoted by equations (4) and (5):
\begin{equation}
S_{out}(t)  = \begin{cases}
 1, V(t) \geq V_{thr} \\
 0, otherwise
\end{cases}
\label{eq:eq4}
\end{equation}

\begin{equation}
V(t) = V_{reset}, when \, S_{out}(t)=1
\label{eq:eq5}
\end{equation}
\end{enumerate}

In this work, a set of 48 \ac{LIF} neurons is employed to convert the 48 channels of processed \ac{sEMG} signals to 48 channels of spike trains. Examples of encoded spike trains for 3 gestures all from electrode 1 are demonstrated in Fig.~\ref{fig:energy-based}a. The parameters in each of the \ac{LIF} neurons were fine-tuned to keep the spike firing rate of each channel under 300 Hz. As the spike trains include precise timing of spikes, higher firing rate could result in saturated neurons which may blur the distinction between significant temporal patterns and lose valuable information. The selection of 300 Hz as the maximum spike firing rate was considered to prevent the network from reaching saturation and causing performance degradation. Each subject shares the same parameters in terms of spike encoding.

\subsection{Network description}

\begin{figure}[!t]
\centerline{\includegraphics[width=\columnwidth]{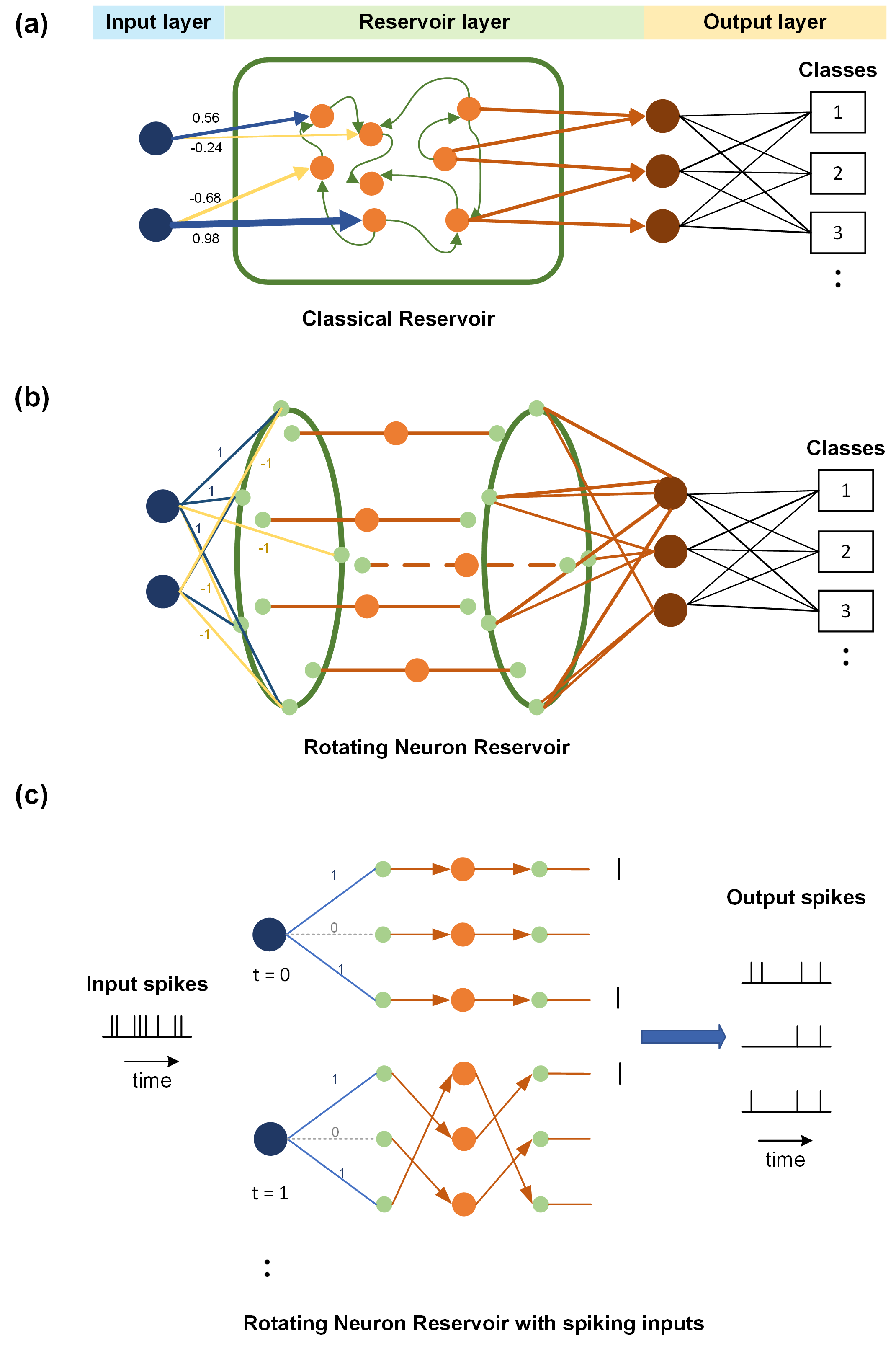}}
\caption {The description of reservoir topologies. The weights of input masks is represented by different colors, dash types and line thicknesses. (a) A classical reservoir topology. The input weight mask $ W_{in}$ is fixed and random from a uniform distribution [-1,1]. In addition, the connections among neurons are also fixed and random. (b)\&(c) An \ac{RNR} topology. A 3D description in (b) and a 2D sketch expanded by time in (c). The connections of the input-to-reservoir layer and reservoir-to-output layer are circularly shifted at each time step. (b) exhibits a binary input mask, randomly selected numbers from \{-1,1\} in conventional \ac{eRNR}. (c) demonstrates a binary input mask \{0,1\} in a three-neuron \ac{sRNR} topology proposed in our work that incorporates spike trains as inputs. The resulting outputs are also spike trains. }
\label{fig:topology}
\end{figure}

A reservoir computing network, as a subset of \acp{RNN}, is particularly suitable for temporal signal processing due to the recurrent connections among its neurons~\cite{lukovsevivcius2009reservoirRNN,lukovsevivcius2012reservoir}. The reservoir projects the inputs to a higher dimension by a set of randomly inter-connected recurrent neurons and generates rich dynamics in the high-dimensional feature space that facilitates linear separability. The values of weights and connections among neurons in the reservoir are fixed, and only the output layer requires training. Compared with other deep \acp{RNN} which require weight updates in every layer with careful selection of learning parameters, the training cost for reservoir computing is lower.

As wearable devices have a strong demand for low-power time series processing, the \ac{PRC} framework was proposed to meet edge computing requirements that incorporate information processing near sensors or into sensors and reduce adaption delay caused by data transmission~\cite{nakajima2020physical}. In our study, we conducted a gesture recognition task, with a preference for processing data proximate to the terminal device that generates it to minimize the need for data transmission to a centralized server, thereby saving computational overhead and decreasing bandwidth usage. Moreover, local data processing mitigates the risk of potential privacy breaches, which is particularly critical in applications involving sensitive patient information. Consequently, we selected a \ac{PRC} architecture called \ac{RNR} to execute the gesture classification task.

In the classical reservoir topology, the weights and connections among neurons are, although fixed, both randomly generated as shown in Fig.~\ref{fig:topology}a, making it hard to implement by hardware components directly. Nevertheless, a simplified cyclic topology was proposed to minimize the complexity of the reservoir without degrading the performance~\cite{rodan2010minimum}. In this structure, the units are organized in a ring topology, and nonzeros elements in reservoir weight matrix $W$ are on the $W_{i,i+1} = 1$ and $W_{1, N} = 1$, described by the matrix in (6):

\begin{equation}
\begin{bmatrix}  0&0&\cdots&0&0&1 \\  1&0&\ddots&0&0&0 \\  0&1&\ddots&0&0&0& \\  \vdots&\vdots&\ddots&\vdots&\vdots&\vdots \\  0&0&\ddots&1&0&0& \\  0&0&\ddots&0&1&0& \\
\end{bmatrix}
\label{eq:eq6}
\end{equation}

This simplified topology was later realized by electronic circuits using rotating elements (analog multiplexers) and known as \ac{eRNR}~\cite{liang2022rotating}. The topology of the\ac{RNR} is demonstrated in Fig.~\ref{fig:topology}b. The connections of the input-to-reservoir layer and reservoir-to-output layer are circularly shifted at each time step. A sketch of the working principle of this circular shift procedure expanded by time is demonstrated in Fig.~\ref{fig:topology}c. The impact of this rotating procedure on the inputs is mathematically the same as the weight matrix $W$ in equation (\ref{eq:eq6})~\cite{liang2022rotating}. The authors conducted comparisons between the \ac{RNR} structure and the single node with delayed feedback structure~\cite{appeltant2011information}, the latter being another widely utilized topology in \ac{PRC}, and claim advantages mainly in terms of (i) lower hardware components costs without \acp{ADC}, (ii) parallel operation decrease system complexity and latency while the delayed feedback line operates in a serial manner, (iii) capable for large-scale integration.

 Based on the description of \ac{RNR} above, we implement it in an \ac{SNN} architecture, and a sketch of the working principle of this \ac{sRNR} is illustrated in Fig.~\ref{fig:topology}c. 

There are three layers in a typical reservoir computing paradigm: an input layer, a reservoir layer, and a readout layer. As mentioned above, the structure of the input is fixed and the reservoir layer is predetermined to rotate periodically at the initialization stage, only the readout layer requires training. Typically, a prevalent way for solving classification tasks using \ac{RC}-related network relies on linear model trained by convex optimization technique like ridge regression, and using a \ac{WTA} algorithm to determine corresponding target class that exceeds a certain threshold~\cite{liang2022rotating,liang2021neuromorphic,zhong2022memristor}. Despite their exceptional training speed, multivariate time series classifiers utilizing a standard \ac{RC} architecture with linear regression fail to attain the accuracy levels achieved by fully trainable neural networks. With the increase of target classes, such as the gesture recognition task in our work that involves 50 distinct gestures, nonlinear methods could demonstrate superior performance. However, the high-dimension and high-precision reservoir states prevent the use of standard nonlinear classifiers being applied on the readout layer, while some dimension reduction procedures on the reservoir states were proposed to apply nonlinear classifiers on the readout layer~\cite{bianchi2020reservoir}. In contrast, we implemented the reservoir in the low-precision spiking domain in this work, solving the challenge of using nonlinear classifiers. Two distinct classifiers were utilized: the \ac{SVM} classifier and a delta learning rule with Softmax activation.

The detailed design of the input and reservoir layers will be introduced as follows and the training of the readout layer will be covered in Section~\ref{ssec:readout}.

\begin{figure}[htp]
\centerline{\includegraphics[width=\columnwidth]{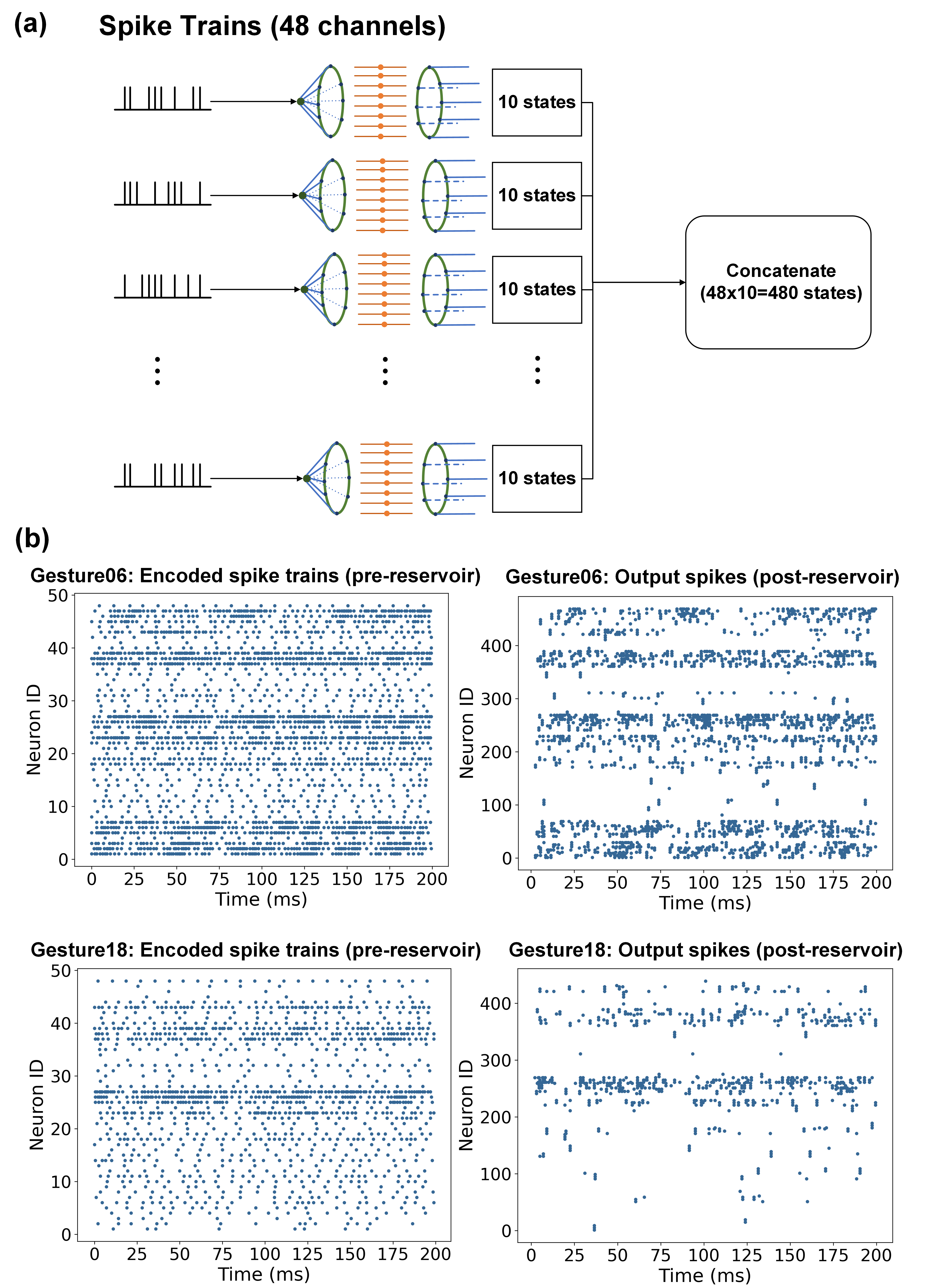}}
\caption {(a) Each 10-neuron reservoir processes an input spike train. The parallel reservoirs project the input spike trains to a higher dimension (from 48 to 480). (b) Two examples of input spike patterns (left) and output spike patterns (right).}
\label{fig:spike patterns}
\end{figure}

\begin{figure}[htp]
    \centering
    \includegraphics[width=0.5\linewidth]{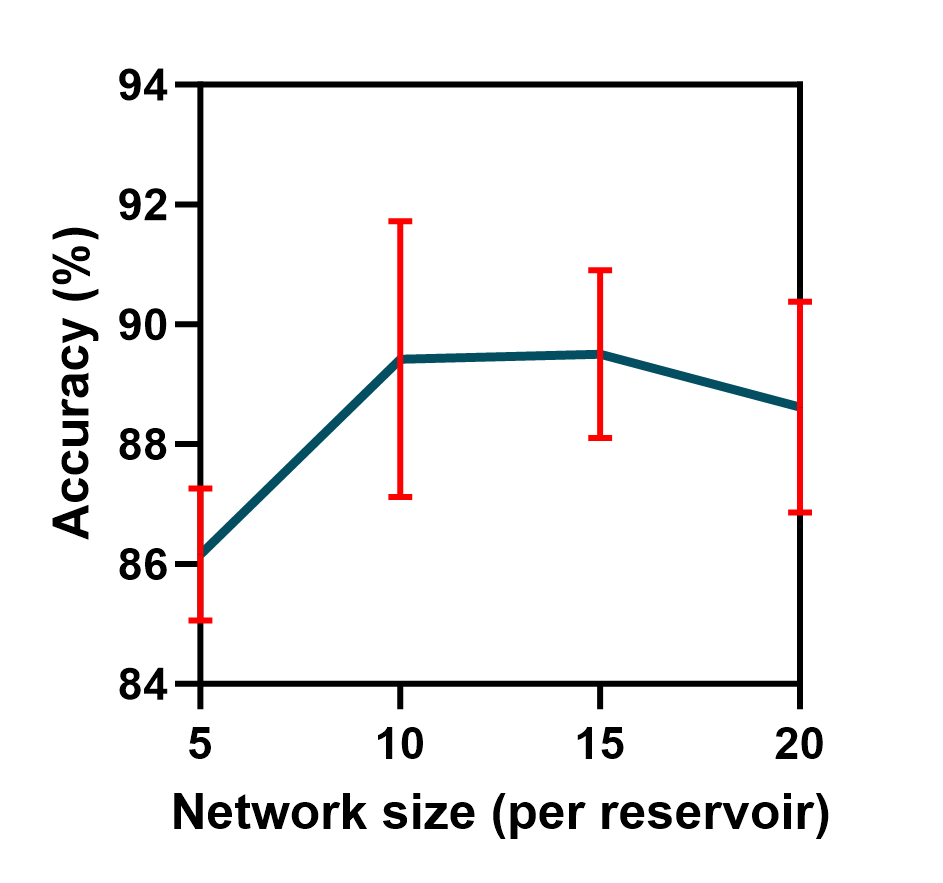}
    \caption{The effect of network size on the classification accuracy of a representative subject by performing exercise B (17 gestures).}
    \label{fig:networksize}
\end{figure}

\begin{enumerate}
\item \textit{Input layer: }The input layer incorporates an input masking procedure that interfaces the input spike trains with the reservoir layer given by equation (7), targeting at increasing state richness~\cite{liang2024physical}: 
\begin{equation}
Spk_{M}^{N \times L}(t) = W_{in}^{N \times 1} \otimes Spk(t)^{1 \times L}
\label{eq:eq7}
\end{equation}
where $Spk(t)$ denotes the input spike trains, $W_{in}$ denotes the input mask matrix, $Spk_{M}(t)$ denotes the masked spike trains, $N$ is the reservoir size (N neurons in the reservoir), and $L$ is the length of the spike train.
The masking matrix $W_{in}$ usually contains randomly selected numbers from a uniform distribution of [-1,1] or randomly chosen binary weights \{-1,1\} in classical \ac{RC} and \ac{eRNR} frameworks respectively as indicated in Fig. ~\ref{fig:topology}a and ~\ref{fig:topology}b. And the 1 and -1 denote the positive/negative signal source in \ac{eRNR}. However, as a spike-based framework proposed in our work, these choices may not be compatible with the network, as negative and floating point numbers should be avoided to add additional hardware costs. At the same time, a sparse connection to establish connection probabilities might be useful in \ac{SNN}~\cite{pyle2017spatiotemporal}. Taking the two factors into account, we finally chose the binary mask \{0,1\} as shown in Fig.~\ref{fig:topology}c. When 0 is applied, it indicates an absence of connection between the input and the \ac{LIF} neuron. Otherwise, it exerts no effect on the connection between the input and \ac{LIF} neuron at that time step.

\begin{figure}[!t]
\centerline{\includegraphics[width=\columnwidth]{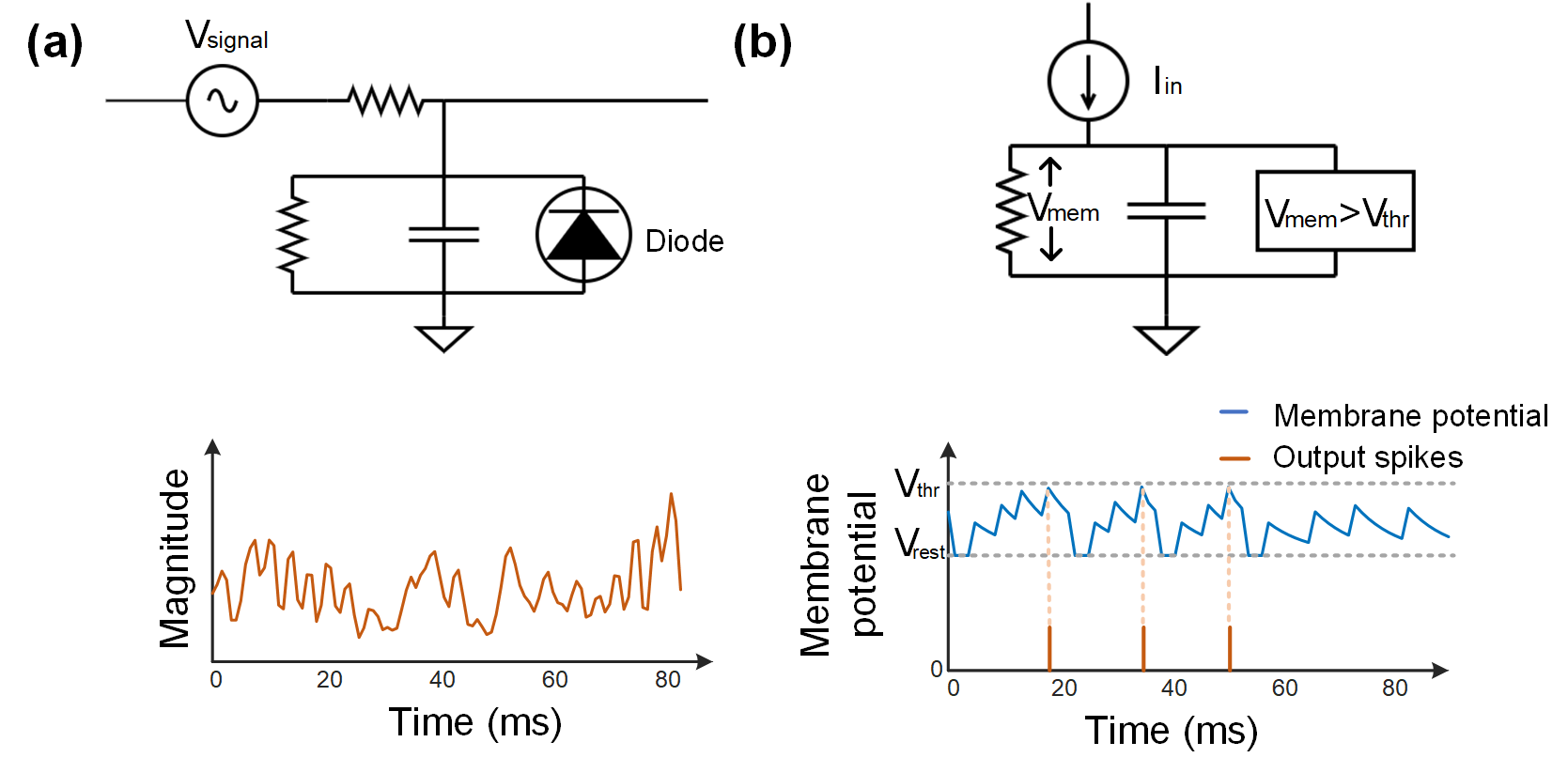}}
\caption {The description of dynamic neurons. (a) The nonlinear integration-\ac{ReLU}-Leakage neuron model and the generated dynamics in terms of continuous output states. (b)The \ac{LIF} neuron model. The neuron will fire a spike when the membrane potential $V_{mem}$ exceeds the threshold $V_{thr}$. Output spikes are considered as the dynamics for training. }
\label{fig:neuron}
\end{figure}

\item \textit{Reservoir layer: }The major difference between \ac{sRNR} and \ac{eRNR} lies in the dynamic neurons in the reservoirs. Fig.~\ref{fig:neuron} provides insights into the circuit for the two neuron models. In \ac{eRNR}, the neuron is a nonlinear integration-\ac{ReLU}-Leakage circuit that performs two important functions: nonlinearity and dynamics. While the nonlinearity is realized by diodes which have similar characteristics with the \ac{ReLU} activiation function and dynamics are provided by the leaky and integrate circuit. However, in \ac{sRNR}, the dynamic neuron is a biologically plausible \ac{LIF} model. It also has integration and leakage characteristics, but different from a nonlinear activation for a continuous analog signal, the neuron receives spikes as inputs and also generates dynamics in the term of spikes. The neuron model we used in our project is called the Lapicque \ac{LIF} neuron model provided by the simulator SNNtorch~\cite{lapicque1907louis,eshraghian2021SNNtorch}. The dynamics in a \ac{LIF} neuron can be quantified by an \ac{ODE} given below which provides a discrete and recurrent representation:

\begin{equation}
\tau\frac{dV_{mem}(t)}{dt} = -V_{mem}(t) + RI_{in}
\label{eq:eq8}
\end{equation}

This model coarsely represents a low-pass filter circuit of a resistor $R$ and a capacitor $C$. The parameter setting is presented in Table~\ref{table 1}. The working principle is the same as discussed in Section~\ref{ssec:encoding}. The neuron will fire a spike when the membrane potential $V_{mem}$ exceeds the threshold $V_{thr}$. Instead of collecting the continuous internal states as shown in Fig.~\ref{fig:neuron}a, we collected the output spikes for training \ac{sRNR}. 

\begin{table}[!t]
\caption{Parameter setting for \ac{LIF} neuron}
\label{table 1}
\centering
\begin{tabular}{|c|c|c|c|}
 \hline
\textbf{Resistance ($R$)} & \textbf{Capacitance ($C$)} & \textbf{Time step } & \textbf{Threshold} \\
\hline
$5$ & $3 \times 10^{-3}$ & $1 \times 10^{-3}$ & $0.5 V$ \\
\hline
\end{tabular}
\end{table}   

The spike trains are injected into the reservoir layer to generate transient dynamics that will be collected for training. There are 48 spike trains and each of the spike trains is injected into a 10-neuron reservoir to form a parallel structure as shown in Fig.~\ref{fig:spike patterns}. The reason for this is that the simulation time for large reservoirs will be extensive and long, a parallel structure allows for neurons states to be computed simultaneously and speeding up the computation times~\cite{vandoorne2011parallel}. The size of 10 was chosen by conducting experiments on a subset of the gestures, as shown in Fig.~\ref{fig:networksize}. The accuracy does not necessarily increase with a larger network size, therefore, a size of 10 was chosen by considering the tradeoff between accuracy and computation cost.
\end{enumerate}
According to the reservoir structure, output spikes have a dimension of 480. An example of input spike trains and output spike trains is shown in Fig.~\ref{fig:spike patterns}b. 

\label{ssec:network}

\subsection{Readout and classification}
\label{ssec:readout}

After the generated output spikes are collected, they will undergo training for classification. As discussed in Section~\ref{ssec:network}, the training takes place in the readout layer only. In our project, we adopted two different supervised classification algorithms to train the readout layer: a \ac{SVM} classifier and a delta rule as the learning rule for the single-layer neural network. The \ac{SVM} classifier is utilized as a baseline to assess the performance of the network and optimize parameters related to spike-encoding and the dynamics generated by reservoirs. Subsequently, a delta rule is applied as it could be fully spiking, offering potential for implementation on neuromorphic chip which has been investigated by several research~\cite{donati2019discrimination,chicca2014neuromorphic,delbrueck1993bump}.

The feature vectors sent to classification consisted of spike counts in a time duration, which refers to a count and binned kernel~\cite{park2013kernel}. In this case, the spiking information is transformed into natural numbers.

\ac{SVM} is a supervised machine learning algorithm used for classification tasks which works by finding hyperplanes that best separate the dataset. In the \ac{SVM} algorithm, we split the training and testing sets in a 4:1 ratio, and the samples for training were shuffled. In addition, the results were cross-validated over 5 different combinations of training and testing sets to evaluate the performance of classification. 

 Delta rule is a foundational learning rule typically applied in the training of a single-layer neural network. The rule adjusts the weights ($w$) of the inputs feature ($x$) based on the difference between the expected output ($y$) and the actual output ($\hat{y}$) to reduce the error in predictions:
 
\begin{equation}
\Delta w = \alpha(y-\hat{y})x
\label{eq:eq9}
\end{equation}

where $\Delta w$ is the weight change and $\alpha$ is the learning rate.

Since it is a multi-class classification task, we applied Softmax activation in the output layer. It converts the linear outputs into probabilities that sum up to 1. The prediction is made by finding the maximum softmax scores (probability). The process is denoted by the following equations: 

\begin{equation}
P(y=i)=\frac{e^{z_{i}}}{\sum^{k-1}_{j=0}e^{z_{j}}}, i\in\{0,...,k-1\}
\label{eq:eq10}
\end{equation}

\begin{equation}
\hat{y} = max(P(y=i),i = 0,...,k-1)
\label{eq:eq11}
\end{equation}
where $P(y=i)$ is the output probability for the $i$-th class, $z_{i}$ is the $i$-th element in the input vector $z$ and $\hat{y}$ is the predicted output class. The \ac{CE} loss is computed at each epoch as the cost function to evaluate how well the model's predictions align with the actual target values.

\section{Analysis and Results}
\label{sec:analysis}

In this section, we presented a \ac{t-SNE} technique which assists in revealing clusters and patterns in the data. Also, we compared the performance by applying different training algorithms. Furthermore, we compared our method with state-of-the-art gesture recognition using \ac{sEMG} signals from Ninapro databases.

\subsection{t-SNE Analysis}

\begin{figure*}[htp]
\centerline{\includegraphics[width=\textwidth]{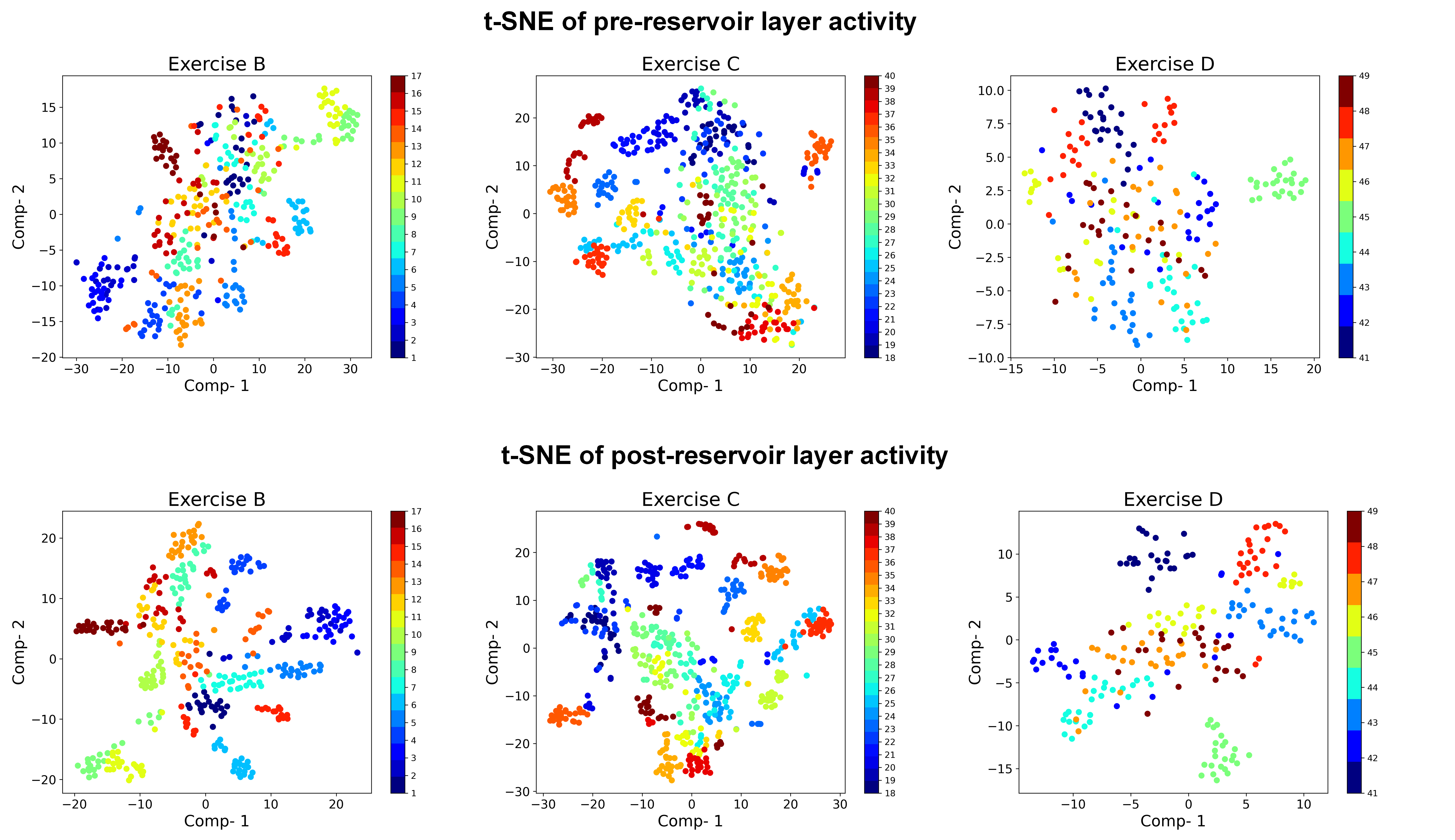}}
\caption {Two-dimension \ac{t-SNE} projections applied on the input layer (pre-reservoir) and output layer(post-reservoir) for Exercises B, C, and D separately in a representative subject. Linear separability is enhanced and observed by more closely clustered patterns following the reservoir layer, where the data are projected into a higher-dimensional feature space.}
\label{fig:tsne}
\end{figure*}

As introduced in previous sections, the reservoir projects the inputs to a higher dimensional feature space to facilitate linear separability. Therefore, to visualize the linear separability, we performed a \ac{t-SNE} analysis on the input layer (pre-reservoir) and output layer (post-reservoir) to observe clusters and patterns inside the data. \ac{t-SNE} was applied separately to exercises B, C, and D described in Ninapro DB2, as visualizing all gestures in the same plot would result in indistinguishable colour discrimination. According to the demonstration in Fig.~\ref{fig:tsne}, projecting the data to a higher-dimensional space facilitates separability since data points belonging to the same class form more closely clustered groups.

\subsection{SVM for Classification}
\label{ssec:SVM}
\acp{SVM} apply kernel functions to map the data into a higher-dimensional space where data are linearly separable. In this work, we applied both a linear kernel and a \ac{RBF} kernel to distinguish all 50 gestures included in Ninapro DB2 and obtained accuracy of $74.6\% \pm 6.3\%$ and $65.2\% \pm 7.4\%$ on the testing sets respectively. Applying a linear kernel achieved around 10\% higher accuracy. This improvement can be attributed to the fact that a linear kernel does not perform any transformation on the data, making it well-suited for data that is already linearly separable. In contrast, the \ac{RBF} kernel, as a nonlinear kernel, projects the data to an infinite-dimensional space which is often used for nonlinear problems. Since the reservoir layer in our proposed method already performs the higher-dimensional mapping which facilitates linear separability, a linear kernel could demonstrate superior classification performance.

\subsection{Delta learning rule for classification}

\begin{figure}[htp]
\centerline{\includegraphics[width=\columnwidth]{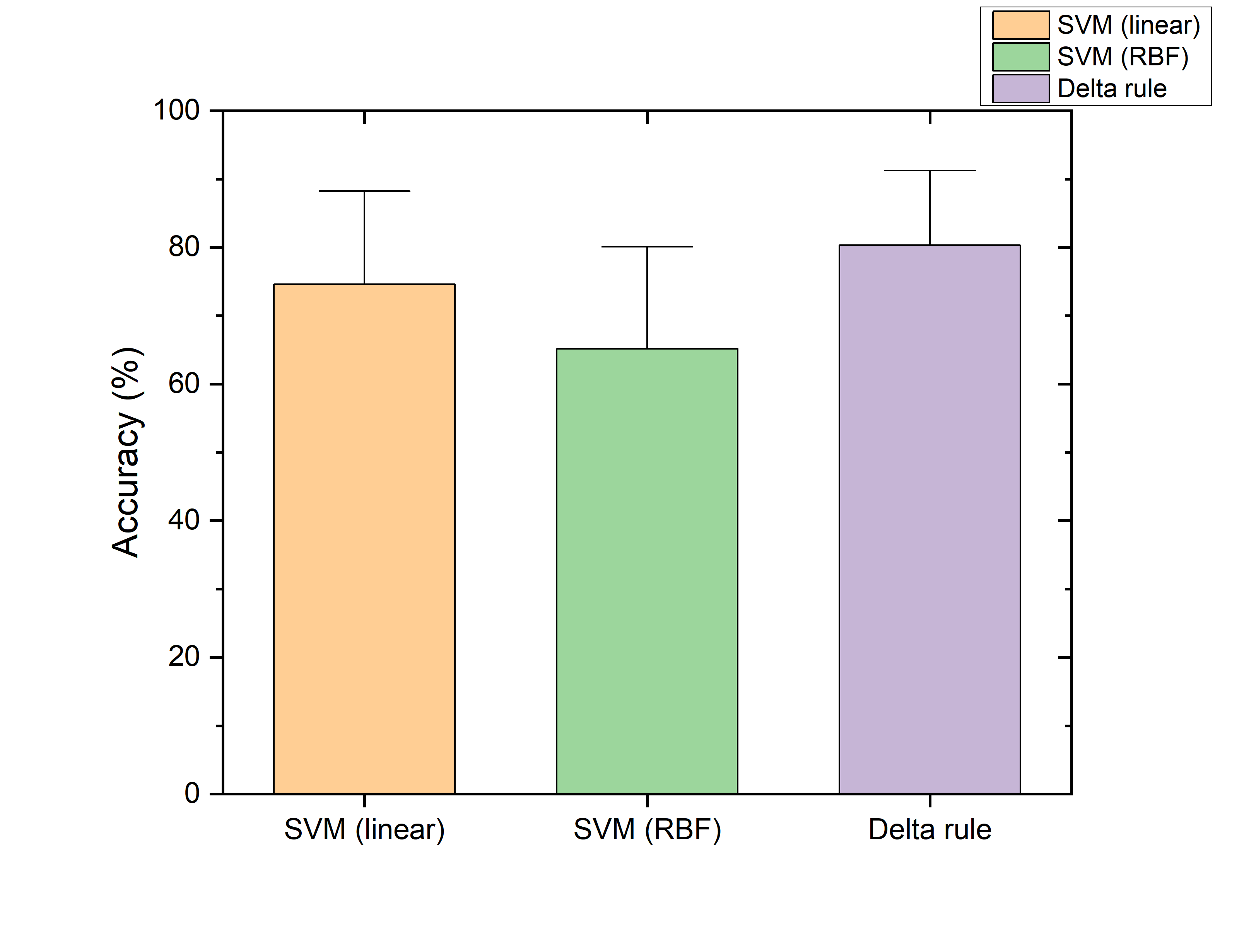}}
\caption {The accuracies for testing sets of using \ac{SVM} (both linear kernel and \ac{RBF} kernel) and delta learning rule with softmax classifier, respectively. Results were averaged over all the subjects with average and standard deviation reported in the bar chart.}
\label{fig:barchart}
\end{figure}

\begin{figure}[htp]
\centerline{\includegraphics[width=\columnwidth]{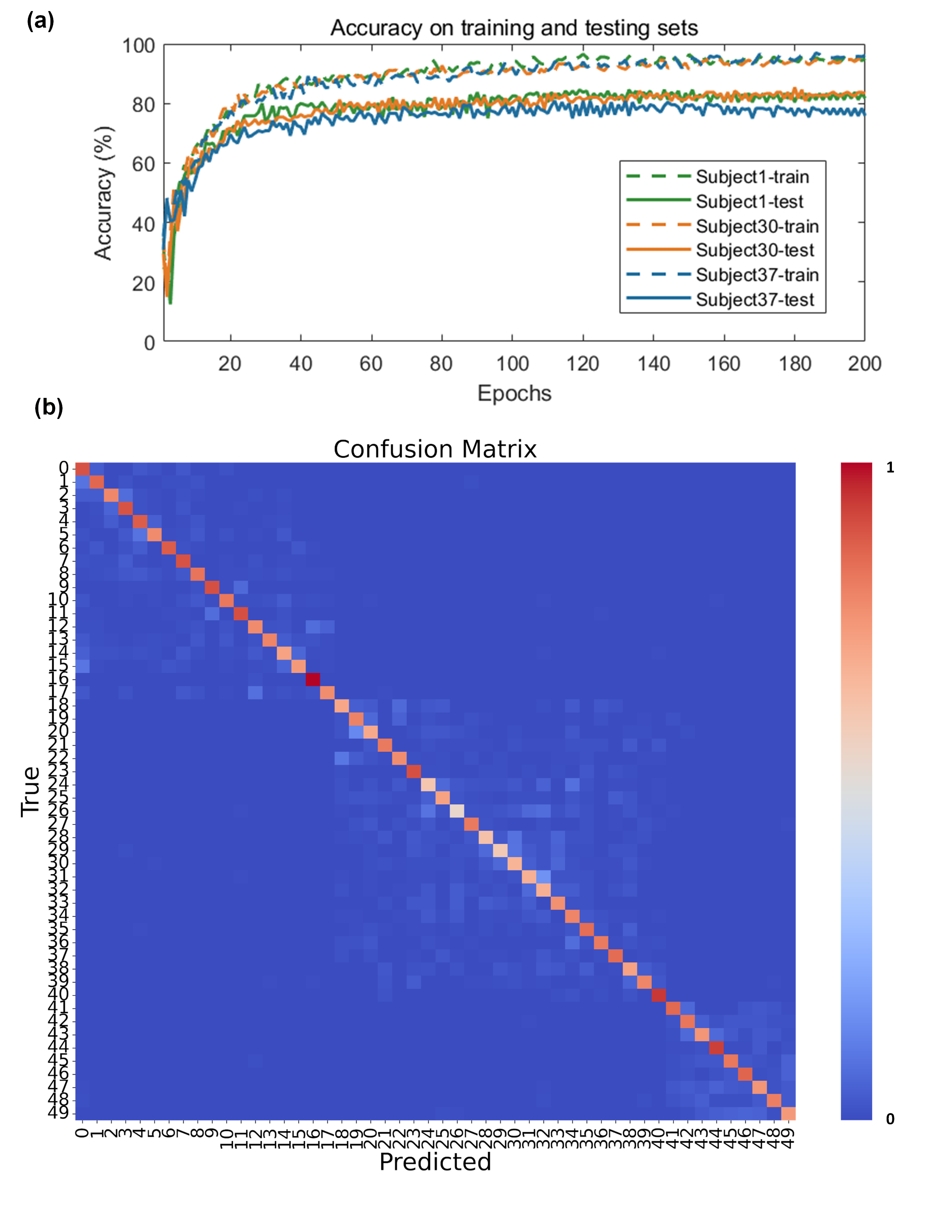}}
\caption {(a) Observed classification accuracies on training sets and testing sets for subject 1, subject 30 and subject 37 through 200 epochs of training. (b) Confusion matrix for classifying 50 gestures on the testing set for all subjects.}
\label{fig:cm&acc}
\end{figure}
Using the learning rule explained in Section~\ref{ssec:readout}, we applied it to classify all 50 gestures by setting the learning rate to 0.005 and batch size to 1. The loss for each subject converges to below 0.2 after 50 epochs of training and achieves a maximum classification accuracy of 94.6\% on the training set and 80.3\% on the testing set after 200 epochs of training by averaging over the 40 subjects. Taking the obtained classification accuracy by \ac{SVM} classifiers as a baseline, we compared the performance of different algorithms as reported in Fig.~\ref{fig:barchart}. Delta learning rule by applying softmax classifier shows a slightly higher classification accuracy. In addition, Fig.~\ref{fig:cm&acc} illustrates the observed accuracies on training and testing sets through training for three representative subjects and a confusion matrix to illustrate the performance of classifying 50 gestures by averaging all subjects on the testing sets using the delta rule.

In addition to the evaluation of classification accuracy, we also assess the model performance based on \ac{PP}, \ac{Sp}, \ac{Se} and F1 score. These metrics can be calculated by using the values of \ac{TP}, \ac{TN}, \ac{FP} and \ac{FN}.

\ac{PP}, also known as precision, is the proportion of predicted positives that are actually positive, defined as follows:
\begin{equation}
PP=\frac{TP}{TP+FP}
\label{eq:eq12}
\end{equation}

\ac{Sp} presents the ability of a classifier to correctly identify all negative instances of a class, defined as follows:
\begin{equation}
Sp=\frac{TN}{TN+FP}
\label{eq:eq13}
\end{equation}

\ac{Se}, also known as recall, is the ability to correctly identify all positive instances of a class, defined as follows:
\begin{equation}
Se=\frac{TP}{TP+FN}
\label{eq:eq14}
\end{equation}

F1 score is the combined assessment of precision and sensitivity, defined as follows:
\begin{equation}
Se=\frac{2\times TP}{2\times TP +FN + FP}
\label{eq:eq15}
\end{equation}
The detailed statistical results of the metrics in percentage mentioned above for each subject are shown in Table~\ref{table 2}, with an overall accuracy of 80.3\%, \ac{PP} of 80.1\%, \ac{Sp} of 99.6\%, \ac{Se} of 77.2\% and F1 score of 78.9\%.

\begin{table}[!t]
\caption{The results of statistical analysis}
\label{table 2}
\centering
\begin{tabular}{c c c c c c }
\hline
\hline
Subject No. & Acc (\%) & PP (\%) & Sp (\%) & Se (\%) & F1 (\%) \\
\hline
1 & 84.6 & 85.3 & 99.6 & 80.1 & 81.5 \\
2 & 80.8 & 79.3 & 99.6 & 77.5 & 80.0 \\
3 & 85.8 & 84.3 & 99.7 & 81.5 & 80.0 \\
4 & 80.4 & 80.5 & 99.6 & 77.7 & 79.1 \\
5 & 83.7 & 85.7 & 99.6 & 82.2 & 83.8 \\
6 & 82.1 & 82.4 & 99.7 & 81.0 & 82.1 \\
7 & 80.4 & 77.4 & 99.5 & 75.3 & 77.0 \\
8 & 79.6 & 80.8 & 99.6 & 78.4 & 80.1 \\
9 & 87.1 & 83.8 & 99.7 & 81.5 & 83.2 \\
10 & 81.7 & 77.7 & 99.6 & 76.4 & 79.5 \\
11 & 80.4 & 77.0 & 99.6 & 74.8 & 77.4 \\
12 & 63.8 & 62.1 & 99.2 & 57.7 & 59.3 \\
13 & 79.2 & 81.3 & 99.6 & 78.9 & 79.7 \\
14 & 73.8 & 74.2 & 99.4 & 70.8 & 72.4 \\
15 & 85.4 & 87.2 & 99.7 & 85.0 & 86.0 \\
16 & 68.3 & 63.0 & 99.3 & 61.9 & 64.8 \\
17 & 87.9 & 87.6 & 99.7 & 83.6 & 85.3 \\
18 & 72.9 & 73.0 & 99.4 & 69.8 & 72.3 \\
19 & 89.2 & 88.6 & 99.8 & 87.8 & 89.1 \\
20 & 71.6 & 74.6 & 99.4 & 72.3 & 73.3 \\
21 & 71.7 & 75.5 & 99.5 & 72.5 & 75.9 \\
22 & 84.2 & 83.9 & 99.7 & 83.3 & 84.9 \\
23 & 91.3 & 91.3 & 99.8 & 89.8 & 90.8 \\
24 & 83.8 & 82.8 & 99.7 & 81.4 & 82.3 \\
25 & 82.1 & 77.1 & 99.5 & 74.6 & 76.6 \\
26 & 82.1 & 83.1 & 99.6 & 79.6 & 80.9 \\
27 & 65.8 & 62.0 & 99.2 & 60.1 & 64.1 \\
28 & 78.8 & 77.6 & 99.5 & 74.5 & 76.8 \\
29 & 77.9 & 81.4 & 99.5 & 76.9 & 79.0 \\
30 & 85.4 & 82.9 & 99.7 & 81.6 & 83.2 \\
31 & 72.1 & 76.6 & 99.4 & 70.1 & 71.2 \\
32 & 86.7 & 87.4 & 99.7 & 84.0 & 84.3 \\
33 & 88.3 & 89.3 & 99.7 & 86.8 & 87.7 \\
34 & 81.7 & 78.2 & 99.6 & 76.1 & 78.5 \\
35 & 86.7 & 86.5 & 99.7 & 83.5 & 84.1 \\
36 & 77.1 & 81.3 & 99.5 & 77.4 & 78.5 \\
37 & 80.8 & 79.9 & 99.5 & 75.8 & 77.9 \\
38 & 82.5 & 79.3 & 99.5 & 76.4 & 79.2 \\
39 & 77.9 & 79.7 & 99.5 & 76.9 & 78.9 \\
40 & 78.3 & 81.3 & 99.5 & 73.0 & 74.6 \\
\hline
overall & 80.3 & 80.1 & 99.6 & 77.2 & 78.9 \\
\hline
\hline
\end{tabular}
\end{table}

\renewcommand{\arraystretch}{1.5}
\begin{table*}[htp]
\caption{Comparison of recent research found in the literature using sEMG-based gesture recognition}
\label{table 3}
\centering
\begin{tabular}{>{\centering\arraybackslash}p{2cm} 
                >{\centering\arraybackslash}p{2cm} 
                >{\centering\arraybackslash}p{1.3cm} 
                >{\centering\arraybackslash}p{1.5cm} 
                >{\centering\arraybackslash}p{1cm} 
                >{\centering\arraybackslash}p{2cm} 
                >{\centering\arraybackslash}p{2cm} 
                >{\centering\arraybackslash}p{2cm}}
\hline
\hline
\textbf{Work} & \textbf{Approach} & \textbf{Feature extraction} & \textbf{Demonstration} & \textbf{\ac{DNN}} & \textbf{Database} & \textbf{No. of Gestures to be classified} & \textbf{Accuracy(\%)}\\ [2pt]
\hline
Atzori \textit{et al.}~\cite{atzori2014electromyography} & RF & Yes& Software & No & Ninapro DB2 & 50 & 75.3 \\  \hline

Fatayer \textit{et al.}~\cite{fatayer2022semg} & ALR-CNN & No &Software & Yes & Ninapro DB2 & 41 & 87.9 \\  \hline

Wei \textit{et al.}~\cite{wei2019surface} & Multi-view CNN & No &Software & Yes & Ninapro DB2 & 50 & 83.7 \\ \hline 

Vitale \textit{et al.}~\cite{vitale2022neuromorphic} & Spiking FCN & No & Hardware & Yes & Ninapro DB5 & 13 & 74.0 \\ \hline

\multirow{2}{*}{Ma \textit{et al.}~\cite{ma2020emg}} & \multirow{2}{*}{Spiking RNN + STDP} & \multirow{2}{*}{No} & Software & \multirow{2}{*}{No} & \multirow{2}{*}{Ninapro DB2} & \multirow{2}{*}{8} & 57.2 \\ 
 &   &   & Hardware &  &  &  & 55.9 \\ \hline

\textbf{This work} & Spiking \ac{RNR} & No & Software & No & Ninapro DB2 & 50 & 80.3\\ 

\hline
\hline
\end{tabular}
\end{table*}  

\subsection{Comparison with the state-of-the-art}

In recent years, significant advancements have been made in the field of sEMG-based gesture recognition, particularly with the development of state-of-the-art methods such as deep learning algorithms. These approaches have demonstrated notable improvements in classification accuracy. Before deep learning, conventional machine learning algorithms relied on feature extraction in the pre-processing stage. A common machine learning technique selected five feature sets consisting of \ac{RMS}, \ac{TD} statistics, \ac{HIST}, \ac{mDWT} and the normalized combination of all above, and achieved an overall classification accuracy of 75.3\% using \ac{RF} to classify all 50 gestures in the Ninapro DB2 dataset~\cite{atzori2014electromyography}. In contrast, \acp{DNN} can automatically learn from data to extract features while maintaining a high classification accuracy. For example, a multi-view \ac{CNN} method achieved an overall accuracy of 83.7\% in classifying all 50 gestures included in Ninapro DB2 while another \ac{ALR} \ac{CNN} method achieved an overall accuracy of 87.9\% in classifying 41 gestures in Ninapro DB2 (all gestures included in exercise B and C)~\cite{wei2019surface,fatayer2022semg}. However, \ac{DNN}-related methods usually bring increased computational costs, making them unsuitable for the lightweight algorithm requirement of edge devices. In addition, continuous monitoring the muscle activities through wearable devices could generate massive \ac{sEMG} data, resulting in an information bottleneck that challenges data transfer and subsequent post-precessing.

Neuromorphic-based approaches were studied to solve these challenges by bringing analogue and continuous \ac{sEMG} signals to the discrete spiking domain in an event-driven mode. Such methods can process data on the sensor side with reduced computational overhead and latency, providing a new solution to wearable devices. Nevertheless, a significant degradation in classification performance arises in \ac{SNN}-based methods. An accuracy of only 57.2\% was obtained in classifying a subset of 8 selected gestures in Ninapro DB2 by applying a spiking \ac{RNN} with a \ac{STDP} learning rule~\cite{ma2020emg}. In addition, the same network was deployed on a configurable neuromorphic chip and achieved a classification accuracy of 55.9\%, only a small reduction compared with the software simulation result.  An improvement to 74.0\% was made by applying a \ac{sFCN} to classify a subset of 13 gestures in Ninapro DB5, also implemented on a neuromorphic chip~\cite{vitale2022neuromorphic}. However, the results remain less competitive than state-of-the-art machine learning algorithms. In our work, we adopted a novel event-based spike encoding scheme for \ac{sEMG} signals and proposed an \ac{sRNR} which attempted the fusion of \ac{PRC} and \ac{SNN} for the first time and achieved an overall accuracy of 80.3\% by applying a delta learning rule and softmax classifier in classifying all 50 gestures in Ninapro DB2, with few parameters monitoring in training a single-layer readout. This marks a significant improvement over the existing state-of-the-art in terms of \ac{sEMG}-based gesture recognition in the \ac{SNN} domain. Although this work is not implemented on a similar neuromorphic chip since the changing connections, as described in Section~\ref{ssec:network}, are not compatible with neuromorphic chips which require initialization of network topologies during the startup phase, it still paves way to the integration into chip level with dedicated hardware design for future research. A summary of the comparison of the state-of-the-art is demonstrated in Table~\ref{table 3}.

\section{Discussion and Conclusion}
In this paper, we proposed a novel \ac{sEMG}-based gesture recognition framework by implementing a \ac{PRC} topology within an \ac{SNN} architecture, alongside an innovative event-based spiking encoding strategy for the \ac{sEMG} signals. We performed a thorough investigation into the proposed approach by a \ac{t-SNE} visualization of internal network activity and two different classifiers to evaluate the classification performance. The results indicate that our approach significantly outperforms existing \ac{SNN}-based methods for the large-scale public dataset with higher classification accuracy, and at the same time, has a lower training cost compared with deep learning algorithms. Statistical evaluations demonstrate that our proposed solution pipelines a new insight into processing real-time signals at the edge for wearable devices, promising compact and lightweight electronic systems for temporal signal processing in wearable devices.

\section*{Acknowledgment}

The authors would like to thank Institute of Neuroinformatics (INI) for generously offering the opportunity to visit and collaborate during this research. The access to resources, expert insights, and professional support provided by INI and its members were invaluable in advancing our work.

\bibliography{reference}

\end{document}